\documentclass[twocolumn,nofootinbib,floatfix,superscriptaddress]{revtex4}        
\usepackage{graphicx}
\usepackage{epsfig}
\usepackage{bm}
\usepackage[T1]{fontenc}
\usepackage[latin9]{inputenc}
\usepackage{amssymb}
\usepackage{float}
\usepackage{amsmath}
\usepackage{dcolumn}
\usepackage{cancel}
\usepackage[colorlinks]{hyperref}
\usepackage[usenames,dvipsnames]{color}
\hypersetup{
	breaklinks=true,
	pdfstartview={FitH},    
	colorlinks=true,       
	linkcolor=blue,          
	citecolor=red,        
	filecolor=magenta,      
	urlcolor=blue,           
	anchorcolor=green,      
	linktocpage=true
}

\newcommand{\be}{\begin{equation}}
\newcommand{\ee}{\end{equation}}

\begin{document}

\title{Universal thermodynamic topological classes of the charged dRGT black string}
\author{Hao Chen}
\email{haochen19@zync.edu.cn}
\affiliation{School of Physics and Electronic Science, Zunyi Normal University, Zunyi 563006,People's Republic of China}
\affiliation{Lanzhou Center for Theoretical Physics, Key Laboratory of Theoretical Physics of Gansu Province, Key Laboratory of Quantum Theory and Applications of MoE, Lanzhou University, Lanzhou 730000,  People's Republic of China}
\author{Meng-Yao Zhang}
\affiliation{College of Big Date and Intelligent Engineering, Guizhou University of Commerce, Guiyang, 550014, People's Republic of China}
\author{Yi Yang}
\affiliation{School of Mathematics and Statistics, Guizhou University of Finance and Economics, Guiyang, 550025, China}

\author{Qihong Huang}
\affiliation{School of Physics and Electronic Science, Zunyi Normal University, Zunyi 563006,People's Republic of China}
\author{Zheng-Wen Long}
\affiliation{College of Physics, Guizhou University, Guiyang 550025, Guizhou, People's Republic of China}
\begin{abstract}
In this study, we explore universal thermodynamic topological classes of charged dRGT black string within both the canonical ensemble and grand canonical ensemble frameworks, and further analyze its asymptotic behavior under limiting parameter regimes. We demonstrate that, while the outermost large black string branch remains thermodynamically stable in both ensembles, the innermost small black string branch exhibits distinctly different stability properties: it is stable in the canonical ensemble but becomes thermodynamically unstable in the grand canonical ensemble, corresponding to the $W^{1+}$ and $W^{0-}$ topological categories, respectively. Furthermore, the local thermodynamic stability of the charged dRGT black string is investigated through the behavior of the heat capacity. These findings demonstrate that the selection of thermodynamic ensemble has a significant influence on the thermodynamic configuration of the charged dRGT black string. In the limit where gravitational effects are neglected, the charge contribution does not modify the underlying topological classification. This implies that the coupling between dRGT massive gravity and the electromagnetic sector is essential for the emergence of nontrivial thermodynamic topology.  These results contribute to a deeper understanding of the black string thermodynamics and provide a novel theoretical basis for exploring the basic properties of quantum gravity.
\end{abstract}

\maketitle

\section{Introduction}
The development of black hole thermodynamics has provided a fundamental theoretical approach for probing the microscopic structure of gravity. Bekenstein first recognized that black holes should possess entropy and formulated the generalized second law of thermodynamics \cite{p1}.  Bardeen et al. established the four laws of black hole thermodynamics \cite{p2}, thereby uncovering the deep connection between spacetime geometry and thermodynamic principles, which offers an essential foundation for investigations of quantum gravity. The subsequent discovery of Hawking radiation further demonstrated that black hole thermodynamics constitutes an important theoretical link among general relativity, quantum mechanics, and thermodynamics \cite{p3}. Among the many intriguing thermodynamic phenomena, A nontrivial phase transition in anti-de Sitter (AdS) spacetime was originally reported by Hawking and Page, occurring between thermal radiation and a thermodynamically stable large Schwarzschild black hole \cite{p4}. Later, this phenomenon was reinterpreted in the context of gauge field theory as a confinement/deconfinement phase transition \cite{p5}. Motivated by the AdS/CFT correspondence \cite{p6}, Kastor et al. developed a systematic formulation of black hole thermodynamics in the extended phase space by treating the cosmological constant as the thermodynamic pressure $P$, with its conjugate quantity naturally identified as the thermodynamic volume $V$. In this framework, the mass of a black hole is reinterpreted not as internal energy but as the enthalpy of the corresponding thermodynamic system \cite{p7}. Such a formulation establishes a unified foundation for analyzing black hole phase transitions and enables a direct thermodynamic analogy between black holes and conventional thermodynamic systems. In recent developments, Xiao et al.generalized the Iyer-Wald formalism to derive the extended black hole thermodynamics naturally \cite{p7.1}. Within this framework, a variety of novel phase transitions and rich thermodynamic structures have been uncovered, attracting extensive investigation \cite{p8,p9,p10,p11,p12,p13,p14,p15,p16,p17,p18,p19,p20,p21,p22}. Nevertheless, despite these important developments, a universal theoretical framework capable of characterizing the intrinsic properties of different black hole systems has not yet been fully established.

Despite the substantial advances achieved in black hole thermodynamics over recent years, constructing a unified description of its universal properties and underlying structures remains a difficult problem. By employing Duan's mapping topological currents theory \cite{p23,p24}, the topological approach offers an alternative framework for investigating black hole thermodynamic behavior. Within this formalism, Wei et al. treated black hole states as topological defects embedded in the thermodynamic parameter space and introduced the associated topological numbers through topological charges, thereby classifying various black hole systems into three distinct categories \cite{p25}. Subsequent studies have further generalized the topological analysis approach for describing the critical behavior of black holes. By introducing a temperature-dependent function, a corresponding topological charge can be attributed to each critical point. In this way, critical points associated with black hole phase transitions are classified into two distinct types-namely, traditional critical points and novel critical points-according to their topological charge properties \cite{p26}. Furthermore, the topological number associated with black holes exhibits a close connection to the residue theorem in complex analysis. By establishing a correspondence between the topological configuration around the zero points of the thermodynamic vector field and the residues defined on the complex plane, the topological properties of black holes can be systematically analyzed and verified through the residue theorem \cite{p27}. This approach supplies a more mathematically rigorous foundation for evaluating topological charges and winding numbers, while also exposing the deeper connection between black hole thermodynamic topology and complex geometric theory. Building upon this framework, The thermodynamic topological classification across a variety of black hole systems has been extensively explored (for further details, see \cite{hh1,hh2,hh3,hh4,hh5,hh6,hh7,hh8,hh9,hh10,hh11,hh12,hh13,hh14,hh15,hh16,hh17,hh18,hh19,hh19.1,hh20,hh20.1,hh21,hh22,hh23,hh24,hh25}).

Recently, Wei et al. systematically investigated the thermodynamic stability of black holes in both the innermost/outermost limits and the high-/low-temperature regimes by constructing the asymptotic structure of the thermodynamic vector field, and further put forward a new universal scheme for the topological classification of black holes. By analyzing the asymptotic behavior of four types of black hole temperature in four-dimensional space-time, it is divided into $W^{1-}$ [Schwarzschild], $W^{0+}$ [Reissner-Nordstr\"om (RN)], $W^{0-}$ [Schwarzschild-AdS], $W^{1+}$ [RN-AdS] thermodynamic topological classes \cite{us1}. Based on this framework, other black hole configurations can also be mapped onto these established types through their asymptotic behavior analysis \cite{us2,us3,us4,us5,us6}. However, the present classification scheme remains relatively restricted and cannot fully encompass all known black hole solutions. In particular, in gauge supergravity, multi-spin Kerr-AdS systems, and certain modified gravity theories, black holes exhibit anomalous thermodynamic stability behaviors in the low- or high-temperature regimes, which are not adequately described by the existing classification framework. To address these limitations, Wu et al. further extended the universal topological classification by introducing an additional topological category along with two new sub-classes on the basis of the original structure \cite{us7,us8}. Moreover, our recent work within the framework of Ho$\breve{r}$ava-Lifshitz gravity shows that the Hawking temperature of charged static black holes exhibits an unconventional stability structure that has not been previously reported. Motivated by this result, a new topological subclass is introduced, which further broadens the theoretical scope of the universal thermodynamic topological categorization of black holes \cite{us9}.

Although the thermodynamic topology of black holes has been extensively explored in a variety of gravitational frameworks, including the Lovelock gravity \cite{hh19.1}, the de Rham-Gabadadze-Tolley (dRGT) massive gravity \cite{hh20},the Gauss-Bonnet gravity \cite{hh20.1}, and the Ho$\breve{r}$ava-Lifshitz \cite{us9}, the corresponding thermodynamic topological properties of black strings have received comparatively little attention. In particular, it remains unclear whether the selection of ensembles can reconstruct the topological configuration of black strings in a manner analogous to black holes, which constitutes an important problem worthy of further investigation. Motivated by this issue, the present work concentrates on the universal thermodynamic topological classification of charged dRGT black string \cite{jj1} within the canonical ensemble and grand canonical ensemble frameworks, with the purpose of clarifying the roles played by ensemble dependence, the dRGT massive gravity, and charge coupling in determining the thermodynamic topological structure of black strings. The organization of this paper is outlined as follows: In Sec. \ref{II}, we provide a concise review of the charged black string solution in dRGT massive gravity. Sec. \ref{III} is devoted to the study of the universal thermodynamic topological classification of charged dRGT black strings within both the canonical and grand canonical ensembles. In Sec.\ref{IIII}, we discuss the thermodynamic topology of charged dRGT black string under three special parameter constraints. The research findings will be summarized in Sec.\ref{IIV}.
\section{The charged dRGT black string solution} \label{II}
We briefly review the theoretical framework of dRGT massive gravity, which is recognized as a nonlinear and ghost-free realization of massive graviton dynamics. By introducing a series of carefully constructed graviton self-interaction terms into the Einstein-Hilbert action, the graviton acquires a finite mass, thereby leading to infrared modifications of Einstein gravity at cosmological scales. Unlike earlier massive gravity models, the dRGT construction consistently eliminates the Boulware-Deser ghost instability \cite{jj2,jj3}, ensuring the internal consistency and stability of the theory. Owing to these advantages, dRGT massive gravity has attracted extensive attention in both cosmological and gravitational studies. In particular, it provides an effective framework for investigating late-time cosmic acceleration, black hole thermodynamics, and possible gravitational mechanisms associated with dark energy phenomena. For a general $d$-dimensional space-time, the action of dRGT massive gravity \cite{jj4,jj5} is written as follows:
\begin{equation}
\mathcal{L}_{\text {massive }}=m_g^2 \sum_i^d c_i \mathcal{U}_i(g, f),
\end{equation}
here, $\mathcal{U}{i}(g,f)$ denotes the graviton interaction potential, $m{g}$ characterizes the graviton mass parameter, while $c_{i}$ represents the corresponding coupling coefficients associated with the massive sector. Throughout this work, we employ the natural unit convention and set the Newtonian gravitational constant to $G=1$. In four-dimensional space-time, the explicit form of the effective graviton potential $\mathcal{U}$ can be written as
\begin{equation}\label{f1}
\mathcal{U}(g, f)=\mathcal{U}_2+\alpha_3 \mathcal{U}_3+\alpha_4 \mathcal{U}_4,
\end{equation}
where $\alpha_{3}$ and $\alpha_{4}$ denote two independent dimensionless constants introduced in the dRGT massive gravity framework. The explicit expressions of the interaction contributions $\mathcal{U}_{2}$, $\mathcal{U}_{3}$, and $\mathcal{U}_{4}$, together with their functional dependence on the metric tensor $g$ and the scalar field sector, are given by
\begin{equation}
\begin{aligned}
& \mathcal{U}_2 \equiv[\mathcal{K}]^2-\left[\mathcal{K}^2\right] \\
& \mathcal{U}_3 \equiv[\mathcal{K}]^3-3[\mathcal{K}]\left[\mathcal{K}^2\right]+2\left[\mathcal{K}^3\right] \\
& \mathcal{U}_4 \equiv[\mathcal{K}]^4-6[\mathcal{K}]^2\left[\mathcal{K}^2\right]+8[\mathcal{K}]\left[\mathcal{K}^3\right]+3\left[\mathcal{K}^2\right]^2-6\left[\mathcal{K}^4\right]
\end{aligned}
\end{equation}
where
\begin{equation}
\mathcal{K}_v^\mu=\delta_v^\mu-\sqrt{g^{\mu v} f_{\mu v}},
\end{equation}
here, the square brackets indicate the trace operation, specifically defined by $[\mathcal{K}]=\mathcal{K}_\mu^\mu$ and $\left[\mathcal{K}^n\right]=\left(\mathcal{K}^n\right)_\mu^\mu$. In a four-dimensional space-time setting, the dRGT massive gravity action coupled with the Maxwell field can be constructed through the inclusion of an additional gauge-field contribution \cite{jj1,jj5}, which is written as
 \begin{equation}
S=\int \mathrm{d}^4 x \sqrt{-g}\left[\frac{1}{2}\left[R+m_g^2 \mathcal{U}\right]-\frac{1}{16 \pi} F_{\mu \nu} F^{\mu \nu}\right],
\end{equation}
where, $R$ denotes the Einstein-Hilbert term, while the electromagnetic field tensor is given by $F_{\mu \nu} \equiv (\nabla_{\mu} A_{\nu} - \nabla_{\nu} A_{\mu})$, with the gauge potential taken in the static form $A_{\mu} = (a(r), 0, 0, 0)$. For convenience, we further perform a parameter redefinition of the graviton potential parameters $\alpha_{3}$ and $\alpha_{4}$ appearing in Eq. (\ref{f1}) by introducing two alternative variables $\beta$ and $\alpha$, given as follows:
\begin{equation}
\alpha_3=\frac{\alpha-1}{3}, \quad \alpha_4=\frac{\beta}{4}+\frac{1-\alpha}{12} .
\end{equation}
The field equations of the theory are derived by performing a variational procedure of the action with respect to the dynamical metric tensor $g_{\mu\nu}$, which yields the following form:
\begin{equation}
G_{\mu \nu}+m_g^2 X_{\mu \nu}=\frac{1}{4 \pi}\left(F_{\mu \alpha} F_\nu^\alpha-\frac{1}{4} g_{\mu \nu} F^2\right),
\end{equation}
here, $X_{\mu \nu}$ represents the term obtained from varying the graviton potential with respect to the metric $g_{\mu \nu}$, and it is commonly referred to as the effective energy-momentum tensor. Its explicit expression of this effective tensor can be expressed as follows:
\begin{equation}
\begin{aligned}
X_{\mu \nu}= & \mathcal{K}_{\mu \nu}-\mathcal{K} g_{\mu \nu}-\alpha\left(\mathcal{K}_{\mu \nu}^2-\mathcal{K} \mathcal{K}_{\mu \nu}+\frac{\mathcal{U}_2}{2} g_{\mu \nu}\right) \\
& +3 \beta\left(\mathcal{K}_{\mu \nu}^3-\mathcal{K} \mathcal{K}_{\mu \nu}^2+\frac{\mathcal{U}_2}{2} \mathcal{K}_{\mu \nu}-\frac{\mathcal{U}_3}{6} g_{\mu \nu}\right).
\end{aligned}
\end{equation}
The Maxwell equations governing the dynamics of the electromagnetic field yield the corresponding field equations, which impose constraints on the radial function $a(r)$, namely, the component of the gauge potential. Specifically, these equations determine the form of $a(r)$ as follows:
\begin{equation}
a(r)=-\frac{\gamma}{\alpha_g r},
\end{equation}
where $\gamma$ denotes an integration constant. To ensure consistency with classical electromagnetic theory in the flat space-time limit, this constant can be expressed through the linear charge density $q$ along the $z$-axis, namely $\gamma^{2}=4q^{2}$. On this basis, within the dRGT massive gravity for a static cylindrically symmetric spacetime, the metric describing the charged black string solution  is constructed by
\begin{equation}
d s^2=-f(r) d t^2+\frac{d r^2}{f(r)}+r^2 d \Omega^2,
\end{equation}
here, $d\Omega^{2}=d\varphi^{2}+\alpha_{g}^{2}dz^{2}$ represents the line element of a two-dimensional (2D) surface, which is consistent with the geometry of the black string configuration. We adopt a cylindrical coordinate system in which the black string exhibits translational symmetry along the $z$-axis \cite{jj1}. The coordinate ranges are given by $-\infty < t < +\infty$, $0 \le r < +\infty$, $-\infty < z < +\infty$, and $0 \le \varphi < 2\pi$. In this coordinate setting, the metric function takes the form as follows:
\begin{equation}\label{xx1}
f(r)=\alpha_m^2 r^2-\frac{4 M}{\alpha_g r}+\frac{4 q^2}{\alpha_g^2 r^2}-\alpha_m^2 c_1 r+\alpha_m^2 c_0
\end{equation}
where
\begin{equation}
\begin{aligned}
\alpha_m^2 & \equiv m_g^2(1+\alpha+\beta), \\
 c_1 &\equiv \frac{h_0(1+2 \alpha+3 \beta)}{1+\alpha+\beta},\\
c_0 & \equiv \frac{h_0^2(\alpha+3 \beta)}{1+\alpha+\beta}.
\end{aligned}
\end{equation}
The metric solution given in Eq. (\ref{xx1}) represents an exact charged black string solution in dRGT massive gravity. In the limiting case $\gamma=0$, it smoothly recovers the neutral black string solution of dRGT massive gravity \cite{jj1}. In addition, by setting $\alpha_{m}=\alpha_{g}$ together with $c_{0}=c_{1}=0$, the solution reduces to Lemos' black string \cite{jj6}. On this basis, within the canonical ensemble, thermodynamic quantities such as the Hawking temperature, entropy, electric potential, and mass can be systematically determined by
\begin{equation}\label{cgc1}
T=\frac{1}{4 \pi r_{h}}\left[\alpha_m^2\left(3 r_{h}^2-2 c_1 r_{h}+c_0\right)-\frac{4 q^2}{\alpha_g^2 r_{h}^2}\right],
\end{equation}
\begin{equation}\label{cc1}
S=\frac{1}{2} \pi \alpha_g r_{h}^2,
\end{equation}
\begin{equation}\label{ch3}
\mu=\frac{2 q}{\alpha_g r_{h}},
\end{equation}
\begin{equation}\label{cc2}
M=\frac{1}{4} \alpha_g \alpha_m^2 r_{h}\left(r_{h}^2-c_1 r_{h}+c_0\right)+\frac{q^2}{\alpha_g r_{h}}.
\end{equation}
Next, we investigate the universal thermodynamic topological classification of the charged dRGT black string within both the canonical ensemble and the grand canonical ensemble frameworks, adopting the dRGT massive gravity parameters $(\alpha_{m}=\alpha_{g}  =1, c_0=4.5, c_1=3)$ consistent with the choice given in Ref. \cite{jj7}.
\section{Universal thermodynamic topological classes of the charged dRGT black string}\label{III}
In the section, we will consider the charged dRGT black string is treated as a topological defect embedded in the thermodynamic parameter space, study its thermodynamic topological classification under the framework of both the canonical ensemble and the grand canonical ensemble, and further explore the influence of charge parameters on the thermodynamic topological structure of the charged dRGT black string. In the canonical ensemble, the charge of the system remains fixed, and the thermodynamic behavior of the black string is mainly determined by the temperature fluctuation. In the grand canonical ensemble, the system maintains a fixed potential, and the charge can be exchanged with the outside world, thus reflecting the effect of the electromagnetic degree of freedom on the thermodynamic topology. Through the comparative study of the two ensembles, the influence of ensemble selection on the thermodynamic stability and universal topological classification of black strings can be further revealed.
\subsection{Topological classes of the charged dRGT black string in the canonical ensemble}
In this subsection, we aim to investigate the universal thermodynamic topological classification of the charged dRGT  black string within the canonical ensemble. Based on the entropy  given in Equation  (\ref{cc1}) and the black string mass presented in Equation  (\ref{cc2}), the generalized off-shell Helmholtz free energy \cite{p25} can be written in the form
\begin{equation}
\begin{aligned}
\mathcal{F} & =M-\frac{S}{\tau} \\
& =\frac{q^2}{r_h \alpha_g}+\frac{1}{4} c_0  \alpha_g \alpha_m^2 r_h+\frac{1}{4}  \alpha_g \alpha_m^2 r_h^3\\
&-\frac{r_h^2 \alpha_g\left(2 \pi+\tau c_1 \alpha_m^2\right)}{4 \tau}.
\end{aligned}
\end{equation}
\begin{figure}[h]
\begin{center}
\includegraphics[width=0.4\textwidth]{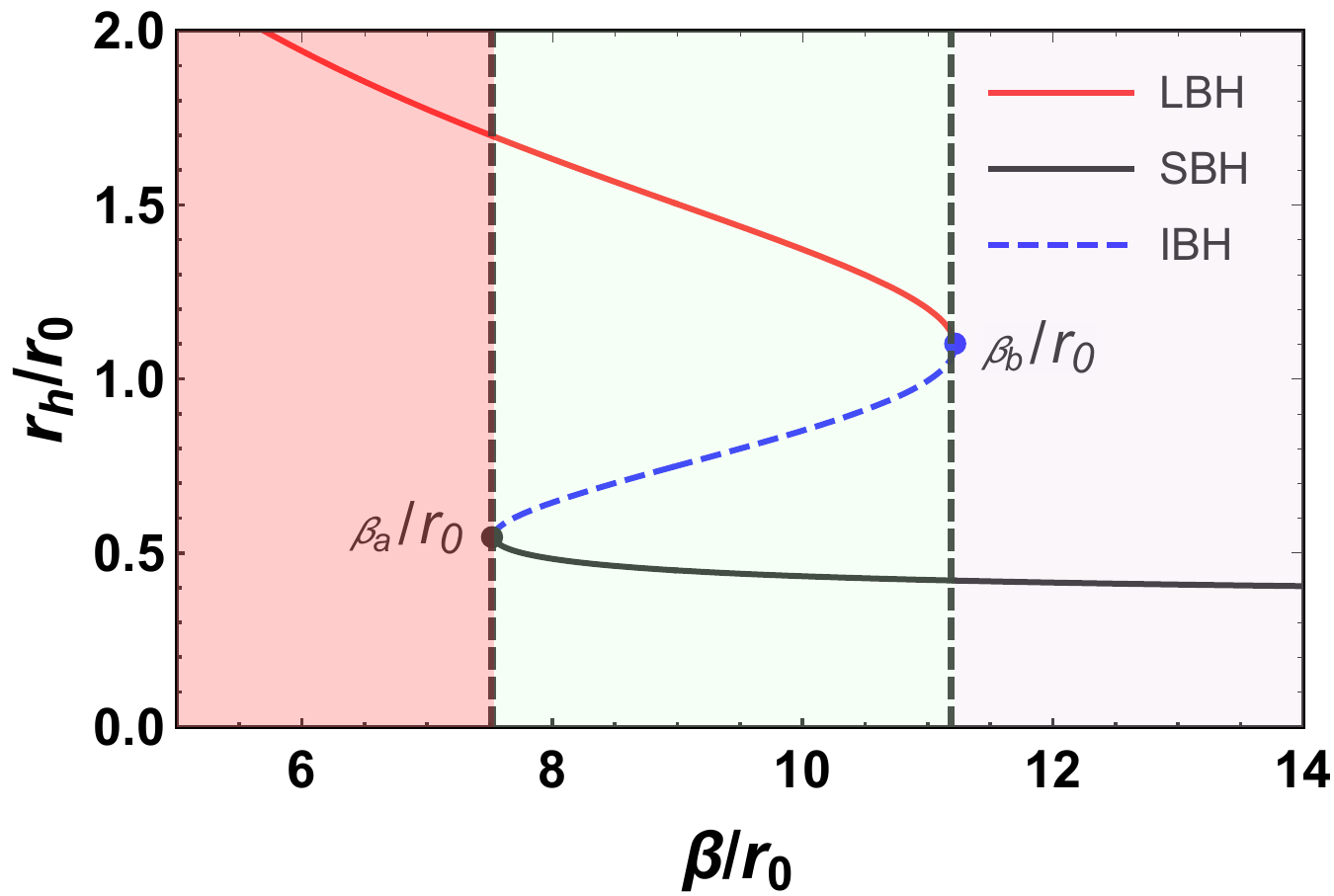}
\caption{In the $r_h-\beta$ plane, the zero points of the vector $\phi^{r_h}$ associated with  the charged dRGT black string in the context of the canonical ensemble, with the parameters $\alpha_{m} / r_0 =\alpha_{g} / r_0 =1$, $q / r_0 = 0.3$, $c_1  / r_0 = 3$ and $c_0  / r_0 = 4.5$. }
\label{m1}	
\end{center}
\end{figure}
\begin{figure}[h]
\begin{center}
\includegraphics[width=0.4\textwidth]{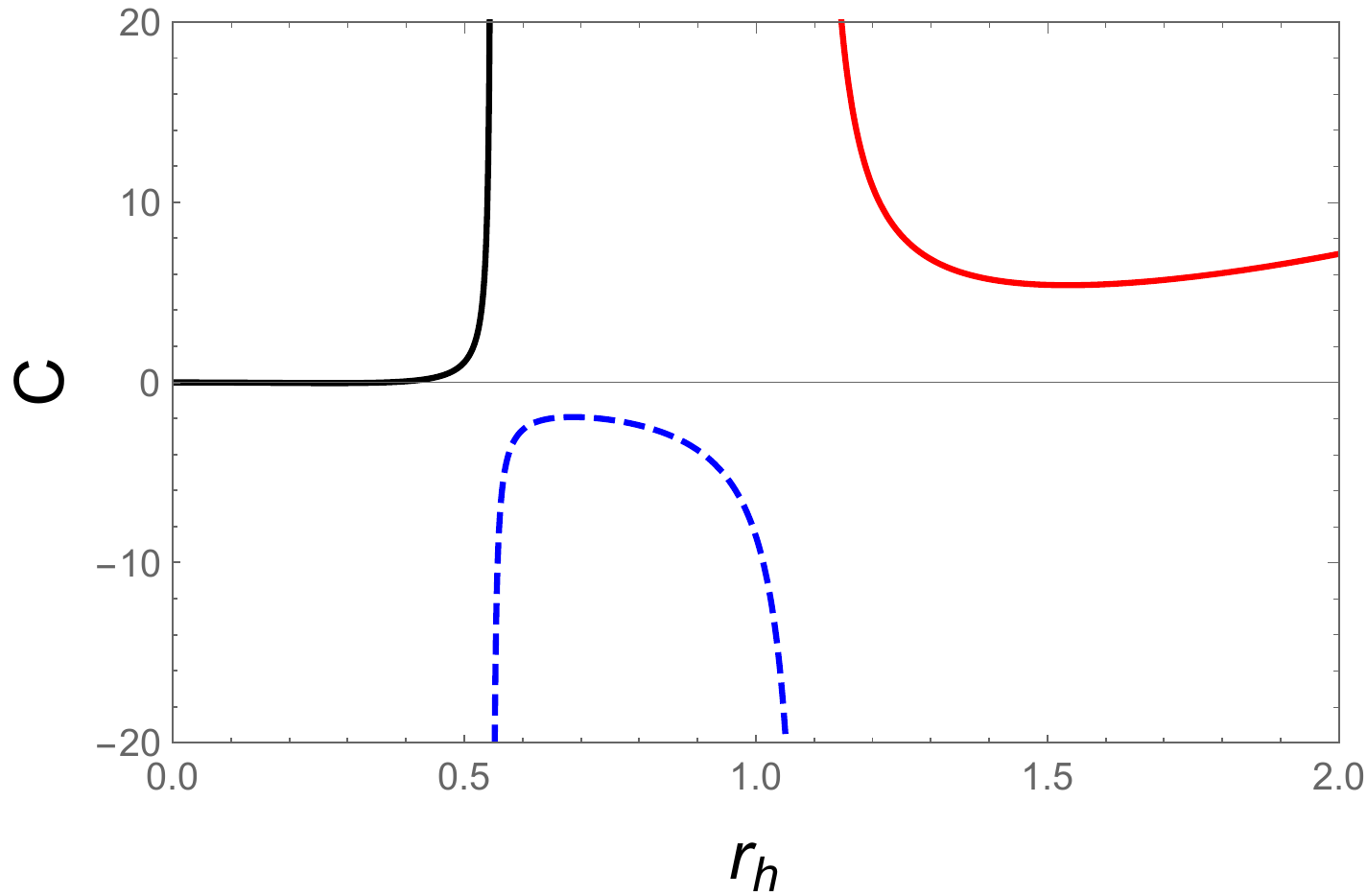}
\caption{Heat capacity $C$ of the the charged dRGT black string as a function of the horizon radius $r_h$ in the canonical ensemble,  with the parameters $\alpha_{m}=\alpha_{g}  =1$, $q  = 0.3$, $c_1  = 3$ and $c_0 = 4.5$.}
\label{my1}	
\end{center}
\end{figure}
\begin{figure}[h]
\begin{center}
\includegraphics[width=0.4\textwidth]{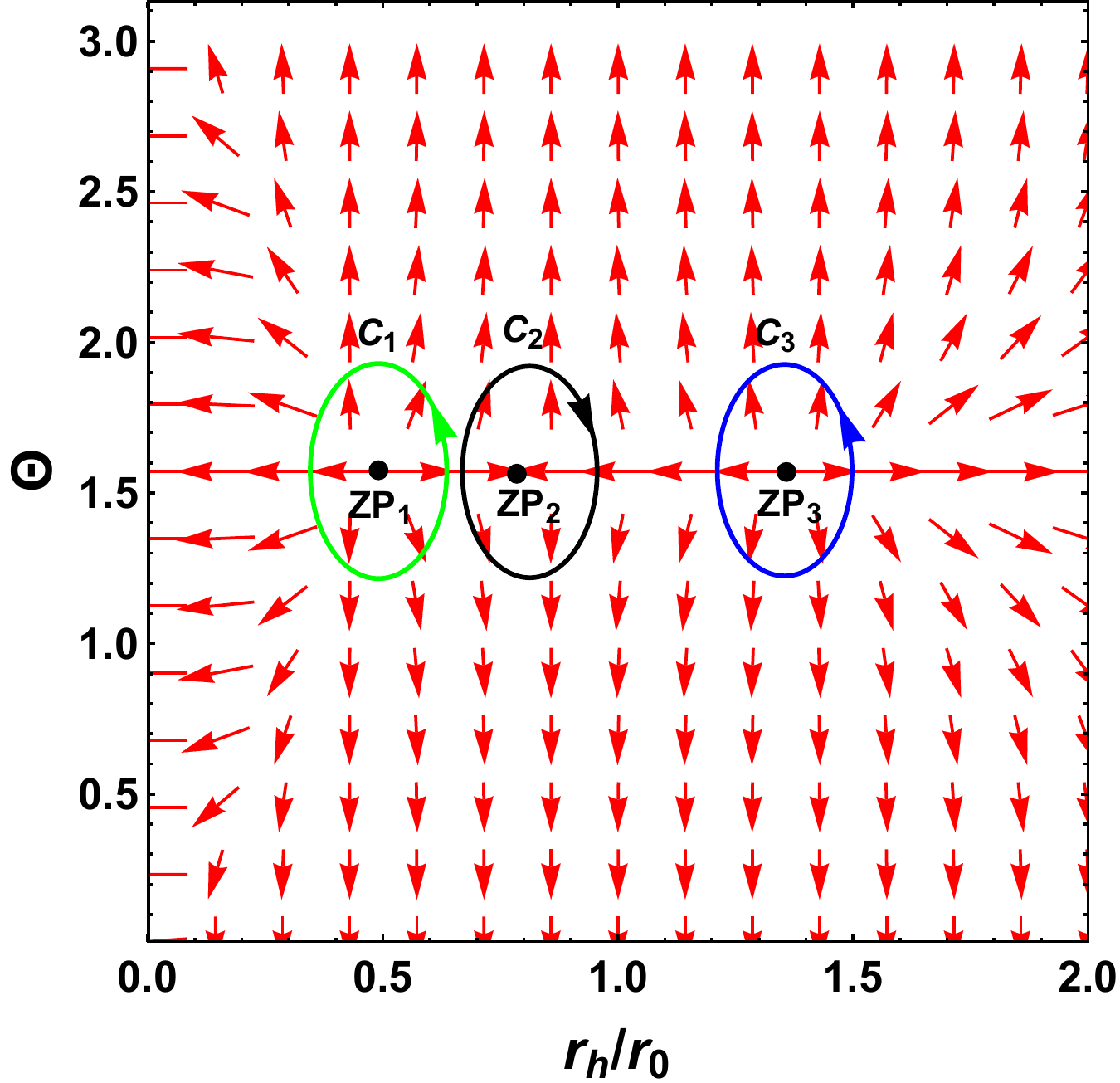}
\caption{The unit vector field $n$ on the $r_{h}-\Theta$ plane for the charged dRGT black string in the context of the canonical ensemble with the parameter  $\beta / r_0 = 10$. }
\label{m2}	
\end{center}
\end{figure}
\begin{figure}[h]
\begin{center}
\includegraphics[width=0.4\textwidth]{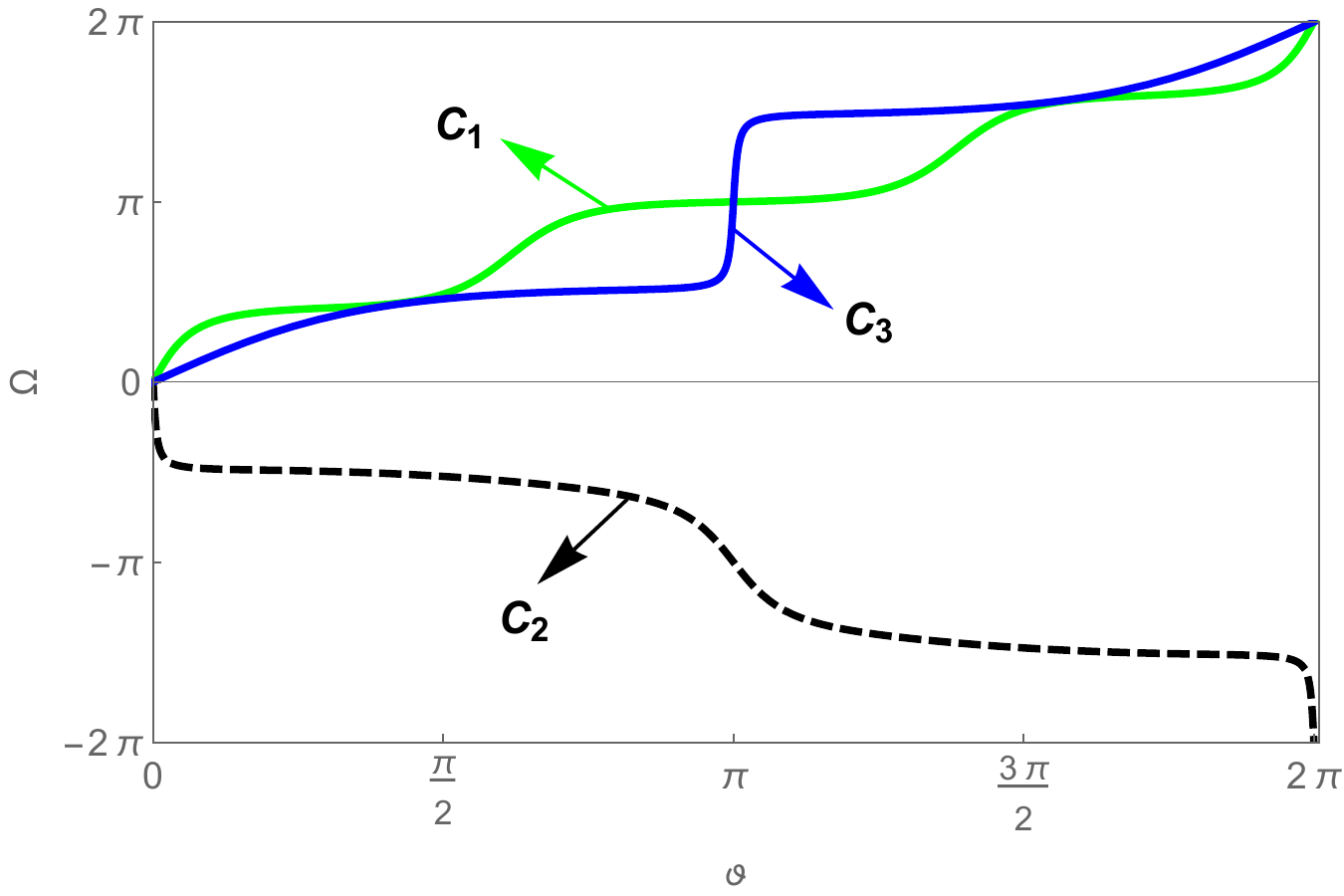}
\caption{$\Omega$ vs $\vartheta$ for the contours $C_1$ (green solid curve), $C_2$ (black dashed curve) and $C_3$ (blue curve), with the parameters $a=b=0.3$.}
\label{m3}	
\end{center}
\end{figure}
In the context of the mapping topological current theory (see Reference \cite{p23,p24}), one may introduce an auxiliary thermodynamic parameter $\tau$, which is physically interpreted as the inverse temperature parameter associated with the thermal cavity surrounding the black hole configuration. In the canonical thermal ensemble, imposing the equilibrium constraint $\tau=\beta=\frac{1}{T}$, where $T$ denotes the Hawking temperature, causes the generalized free-energy to reduce to its on-shell expression. Employing this thermodynamic parametrization, subsequently, a two-component vector field serving as the topological order parameter can be introduced \cite{us1}, whose generic representation is expressed as follows:
\begin{equation}
\begin{aligned}
\phi=\left(\phi^{r_h}, \phi^{\Theta}\right)=\left(\frac{\partial \tilde{\mathcal{F}}}{\partial r_h}, \frac{\partial \tilde{\mathcal{F}}}{\partial \Theta}\right)
\end{aligned}
\end{equation}
where, the function $\tilde{\mathcal{F}}$ is given by
\begin{equation}
\mathcal{F}+\frac{1}{\sin \Theta}
\end{equation}
An auxiliary parameter $\Theta$ is introduced to simplify the analysis of the topological configuration, and its value $(\Theta=\frac{\pi}{2})$ is directly related to the zero-point structure of the vector field.  Within this framework, a winding number can be attributed to each black string configuration, serving to characterize its topological properties and to encode the corresponding topological charge structure. Specifically, locally stable configurations correspond to a winding number $\omega=+1$, whereas locally unstable configurations correspond to $\omega=-1$. By algebraically summing all local winding numbers, one obtains the global topological number. This topological invariant governs the overall topological classification of the black string system as well as its stability properties. Based on this, we can establish two components ($\phi^{r_{h}} $ and $\phi^{\Theta}$)  of the vector field
\begin{equation}
\phi^{r_{h}}=-\frac{q^2}{ \alpha_g r_h^2}+\frac{\alpha_g\left(-4 \pi r_h+\tau\left(c_0+r_h\left(-2 c_1+3 r_h\right)\right) \alpha_m^2\right)}{4 \tau}
\end{equation}
and
\begin{equation}
\phi^{\Theta}=-\cot \Theta \csc \Theta.
\end{equation}
By imposing the condition $\phi^{r_h}=0$, one can derive the following result
\begin{equation}\label{cc3}
\tau=\beta=\frac{4 \pi \alpha_g^2 r_h^3}{c_0 \alpha_g^2 \alpha_m^2 r_h^2-2 c_1 \alpha_g^2 \alpha_m^2 r_h^3+3 \alpha_g^2 \alpha_m^2 r_h^4-4 q^2}.
\end{equation}
It can be directly inferred from Eq. (\ref{cc3}) that, for the charged dRGT black string, the inversion temperature exhibits markedly distinct asymptotic behaviors in different limiting regimes. In particular, under the extremal condition $r_{h} \rightarrow \infty$, the inversion temperature gradually approaches zero. In contrast, when the event horizon radius  $r_{h}$ approaches the the minimal horizon radius $r_{m}$, the inversion temperature increases rapidly and diverges due to the singular behavior of the thermodynamic response function. Therefore, the asymptotic behavior of the inversion temperature can be characterized in the following form
\begin{equation}\label{c4}
\beta\left(r_m\right)=\infty \quad \text { and } \quad \beta(\infty)=0.
\end{equation}
\begin{figure}[h]
\begin{center}
\includegraphics[width=0.4\textwidth]{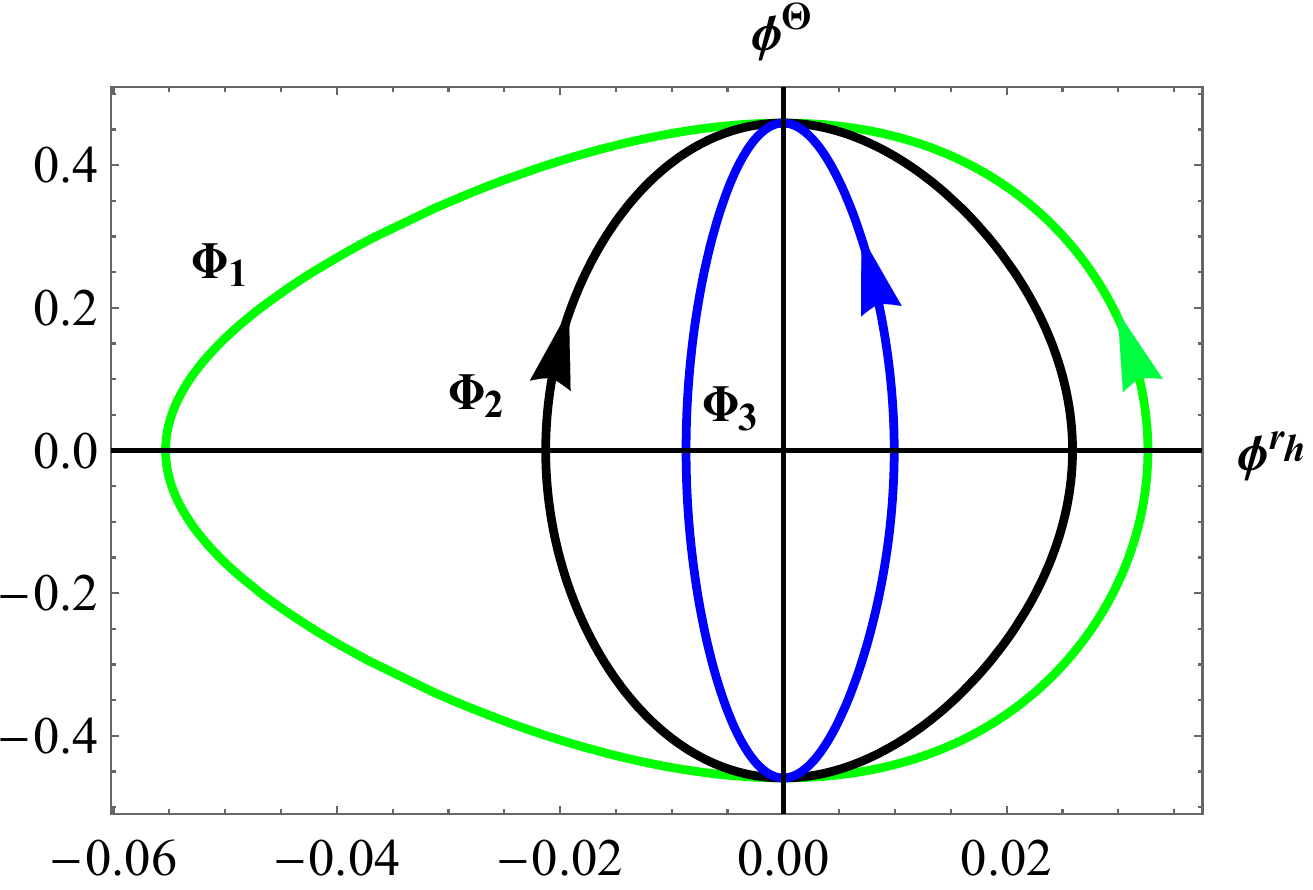}
\caption{The contours $\Phi_i$ illustrate the variations in the components of the vector field $\phi$ as the paths $C_i$ depicted in FIG. 2 are followed for the charged dRGT black string in the context of the canonical ensemble.}
\label{m6}	
\end{center}
\end{figure}
To provide a clearer visualization of the relationship between the inversion temperature $\beta$  and the event horizon radius $r_{h}$,  Fig. (\ref{m1}) depicts the variation of the event horizon radius with respect to the inversion temperature. The thermodynamic states of the charged dRGT black string can be categorized into three separate branches: the small black string branch and the large black string branch, both of which are associated with thermodynamic stability, as well as an intermediate branch that exhibits thermodynamic instability. A more detailed inspection reveals the existence of a black generation point $(\frac{d^2 \beta }{d r_{h}^2}>0)$  as well as a blue annihilation point $(\frac{d^2 \beta }{d r_{h}^2}<0)$, which are responsible for the emergence and disappearance of the branch structure, respectively. In the particular case where $\alpha_{m} / r_0 =\alpha_{g} / r_0 =1$, $q / r_0 = 0.3$, $c_1  / r_0 = 3$ and $c_0  / r_0 = 4.5$, these two characteristic points are located at the critical positions determined by $\frac{\beta_a}{r_0}=7.533$ and $\frac{\beta_b}{r_0}=11.216$, respectively.  Furthermore, as a key physical quantity describing the local thermodynamic behavior of the system, the heat capacity provides an effective measure of the stability of black string configurations. In particular, black string branches with positive (negative) heat capacity correspond to thermodynamically stable (unstable) states. On this basis, the heat capacity can be employed to examine the stability properties of different black string branches across various horizon radius intervals in Fig. (\ref{m1}), thereby serving as a consistency check for the reliability of the results. By combining the Hawking temperature given in Eq. (\ref{cgc1}) with the entropy in Eq. (\ref{cc1}), the heat capacity of the black string in the canonical ensemble can be written as
\begin{equation}
\begin{aligned}
C&=T\left(\frac{\partial S}{\partial T}\right)\\
& =\frac{\pi r_h^2 \alpha_g\left(-4 q^2+r_h^2\left(c_0+r_h\left(-2 c_1+3 r_h\right)\right) \alpha_g^2 \alpha_m^2\right)}{12 q^2+r_h^2\left(-c_0+3 r_h^2\right) \alpha_g^2 \alpha_m^2}.
\end{aligned}
\end{equation}
Fig. (\ref{my1}) illustrates the dependence of the heat capacity on the horizon radius, revealing three distinct black string branches: a stable small black string branch with positive heat capacity $( r_{h} \subset( 0, 0.5477 ) )$, an unstable intermediate black string branch characterized by negative heat capacity $( r_{h} \subset ( 0.5477, 1.0954 ) )$, and a stable large black string branch with positive heat capacity $( r_{h} \subset ( 1.0954, \infty ) )$. These results are consistent with the stability behavior indicated by the inversion temperature curve in Fig. (\ref{m1}), further confirming the validity of employing the inversion temperature parameter to diagnose the thermodynamic stability of black strings. Moreover, the existence of a three-branch structure consisting of stable small, unstable intermediate, and stable large black string states-serves as a key signature of the first-order small/large black string phase transition, analogous to the liquid-gas phase transition in the van der Waals fluid (see Refs. \cite{cp0,cp1} for further discussion). Starting from the critical condition $(\frac{\partial T}{\partial r_h}=\frac{\partial^2 T}{\partial r_h^2}=0)$, one can determine the critical radius $r_{c}=0.866$, the critical charge $q_{c}=0.375$, and the critical temperature $T_{c}=0.392$. When the charge satisfies $q>q_{c}=0.375$, the system admits only a single thermodynamically stable branch. As a result, the phase structure of the charged dRGT black string becomes single-valued, and no phase transition behavior occurs. In addition,  Fig. (\ref{m2}) presents the zero points distribution of the unit vector field. When the inversion temperature  parameter satisfy the condition $(\beta / r_0 = 10)$, three zero points appear at $(0.4329, \frac{\pi}{2})$, $(0.8511, \frac{\pi}{2})$, and $(1.3722, \frac{\pi}{2})$, respectively. In order to further evaluate the winding number associated with each zero point, three  contours  $(C_1)$, $(C_2)$, and $(C_3)$ are introduced. Based on the general parameterization scheme and the definition of the deflection angle given in Ref. \cite{p26}, the topological charges associated with the three contours  shown in Fig. (\ref{m3}) can be determined, yielding values of $Q_{ZP_{1}}=1$, $Q_{ZP_{2}}=-1$, and $Q_{ZP_{3}}=1$, respectively. We further examine the asymptotic properties of the vector field $\varphi$ in the vicinity of the boundary specified by Eq. (\ref{c4}). The boundary structure is represented by the closed contour $(C=I_{1}\cup I_{2}\cup I_{3}\cup I_{4})$. Owing to the definition of $\varphi$ , the vector field remains perpendicular to the segments $(I_{2})$ and $(I_{4})$ \cite{us1}, implying that the dominant asymptotic features are governed by its behavior along $(I_{1})$ and $(I_{3})$. The vector field tends to orient either leftward or rightward as $(r_{h}\rightarrow r_{m})$ or $(r_{h}\rightarrow \infty)$, respectively, with the orientation determined by the component value of the vector field. The associated directional configurations are listed in Table (\ref{tb1}). ig. (\ref{m6}) clearly demonstrates the directional evolution behavior of the vector field along the chosen contours. By analyzing the clockwise and anticlockwise orientations of the corresponding closed loops, the winding numbers associated with the zero points $(ZP_{1}, ZP_{2}, ZP_{3} )$  can also be determined as $+1$, $-1$, and $+1$, respectively. The universal thermodynamic behavior of the charged dRGT black string within the canonical ensemble can be summarized as follows: in the inverse temperature limits $(\beta \rightarrow \infty)$ and $(\beta \rightarrow 0)$, the system corresponds respectively to the low-temperature and high-temperature regimes, and its thermodynamic configuration consistently exhibits a stable black string phase structure. This behavior implies that the charged dRGT black string and the charged RN-AdS black hole have the same topological structure properties, and both belong to the $(W^{1+})$ class \cite{us1}.

\begin{table*}[ht]
    \caption{The orientation of the charged dRGT black string is determined by the direction of the arrow associated with $\phi^{r_{h}}$, which corresponds to the relevant topological number.}
    \centering
    \begin{tabular}{|l|c|c|c|c|c|}
        \hline
        \textbf{Black hole solutions }& ${I_1}$ & $I_2$ & ${I_3}$ & ${I_4}$ & ${W}$ \\
        \hline
        the charged dRGT black string in canonical ensemble  & $\rightarrow$ & $\uparrow$ & $\leftarrow$ & $\downarrow$ & 1 \\
        \hline
         the charged dRGT black string in grand  canonical ensemble & $\rightarrow$ & $\uparrow$ & $\rightarrow$ & $\downarrow$ & 0 \\
        \hline
    \end{tabular}
   \label{tb1}
\end{table*}
\subsection{Topological classes of the charged dRGT black string in the grand canonical ensemble}
\begin{figure}[h]
\begin{center}
\includegraphics[width=0.4\textwidth]{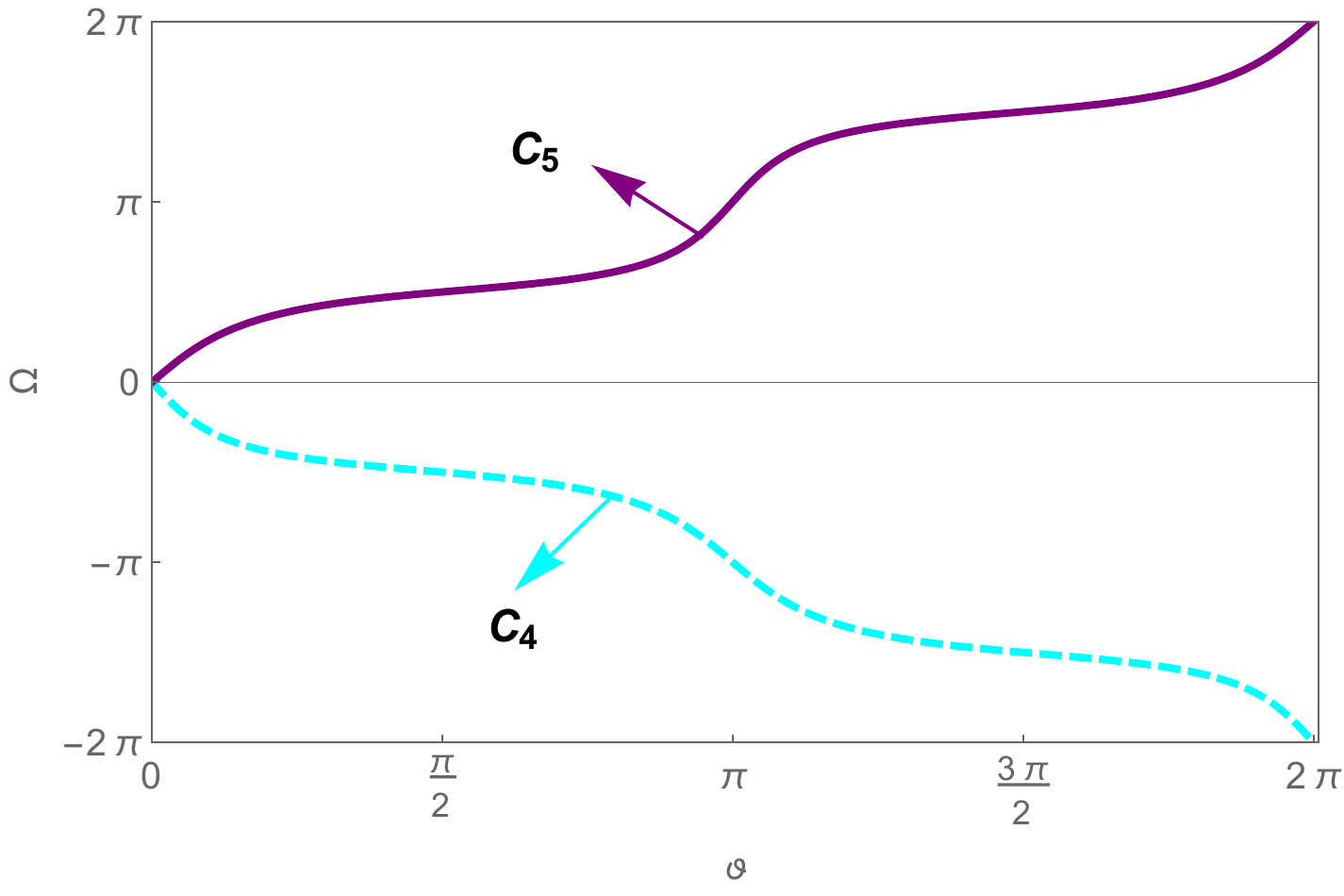}
\caption{$\Omega$ vs $\vartheta$ for the contours $C_4$ (cyan dashed curve) and $C_5$  (purple curve), with the parameters $a=0.15$ and $b=0.4$. }
\label{m9}	
\end{center}
\end{figure}

\begin{figure}[h]
\begin{center}
\includegraphics[width=0.4\textwidth]{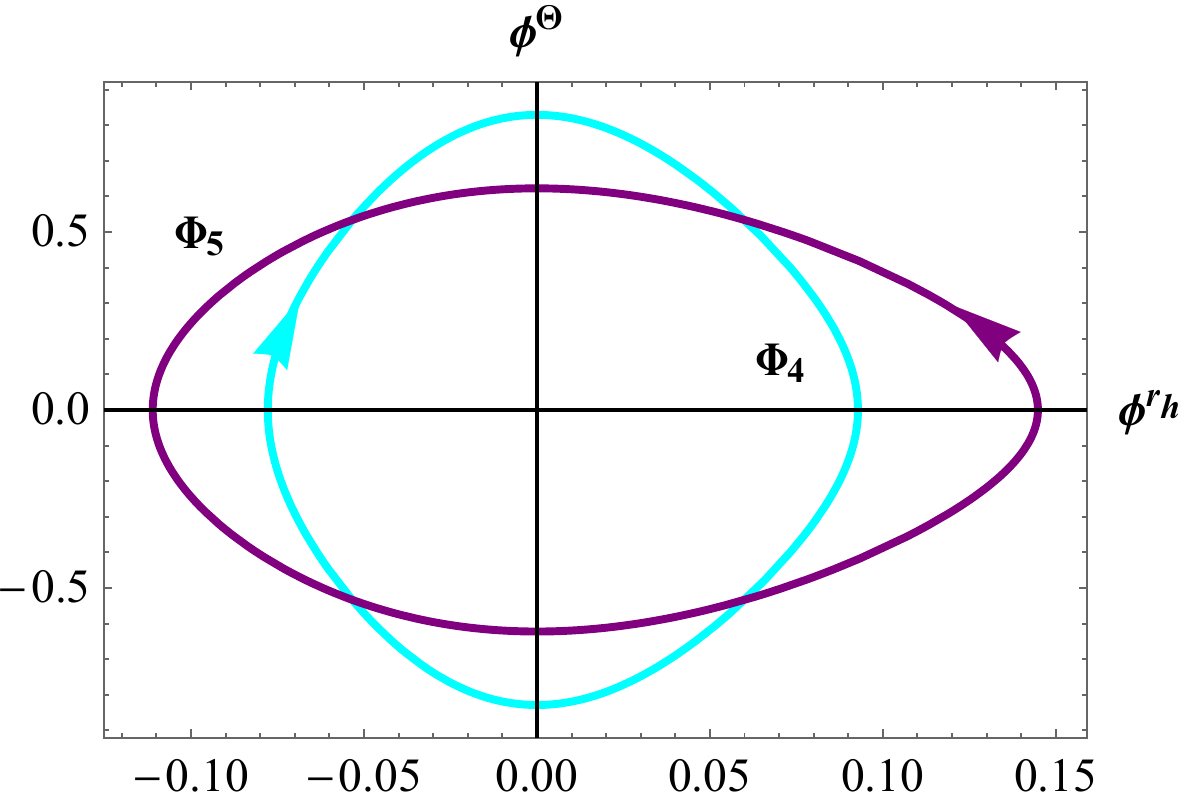}
\caption{The contours $\Phi_i$ illustrate the variations in the components of the vector field $\phi$ as the paths $C_i$ depicted in FIG. 6 are followed for the charged dRGT black string in the context of the grand canonical ensemble. }
\label{m10}	
\end{center}
\end{figure}

In this subsection, we aim to investigate the universal thermodynamic topological categorization of the charged dRGT black string within the grand canonical ensemble. Under this setup, the system is regarded as an open configuration in which charge can be exchanged with the surroundings through processes of transfer, creation, or annihilation, and the variation of charge is controlled by the chemical potential given in (\ref{ch3}). Thus, the generalized off-shell Helmholtz free energy, expressed by the chemical potential in place of the charge, is given by
\begin{equation}
\begin{aligned}
\mathcal{F} & =M-\mu Q -\frac{S}{\tau} \\
& =\frac{\left(-u^2 \tau+\tau c_0 \alpha_m^2+\tau \alpha_m^2 r_h^2-\left(2 \pi+\tau c_1 \alpha_m^2\right) r_h\right) \alpha_g r_h}{4 \tau}.
\end{aligned}
\end{equation}
\begin{figure}[h]
\begin{center}
\includegraphics[width=0.4\textwidth]{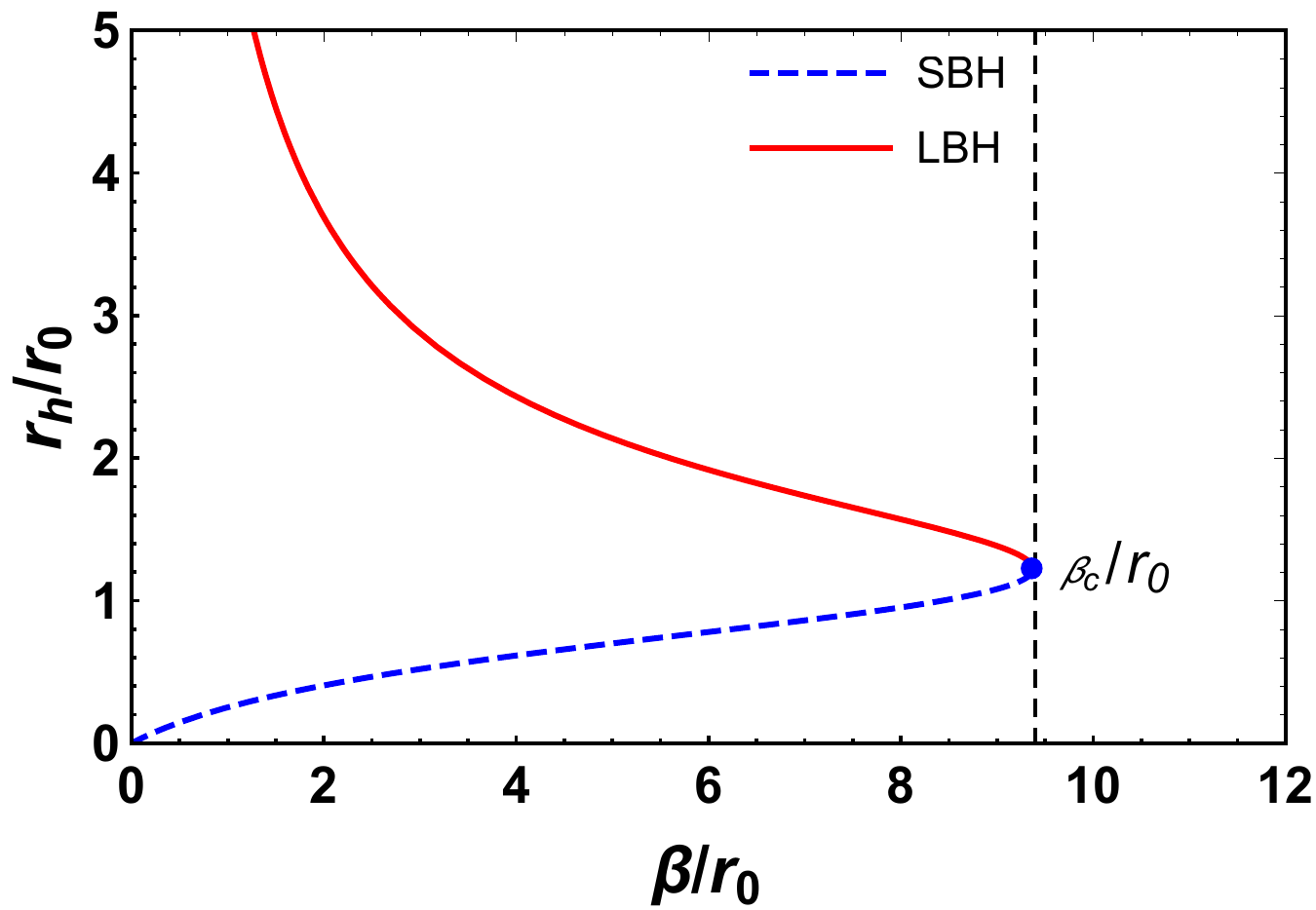}
\caption{In the $r_h-\beta$ plane, the zero points of the vector $\phi^{r_h}$ associated with the charged dRGT black string in the context of the grand canonical ensemble, with the parameters $\alpha_{m} / r_0 =\alpha_{g} / r_0 =1$, $u / r_0 = 1$, $c_1  / r_0 = 3$ and $c_0  / r_0 = 4.5$. }
\label{m7}	
\end{center}
\end{figure}
\begin{figure}[h]
\begin{center}
\includegraphics[width=0.4\textwidth]{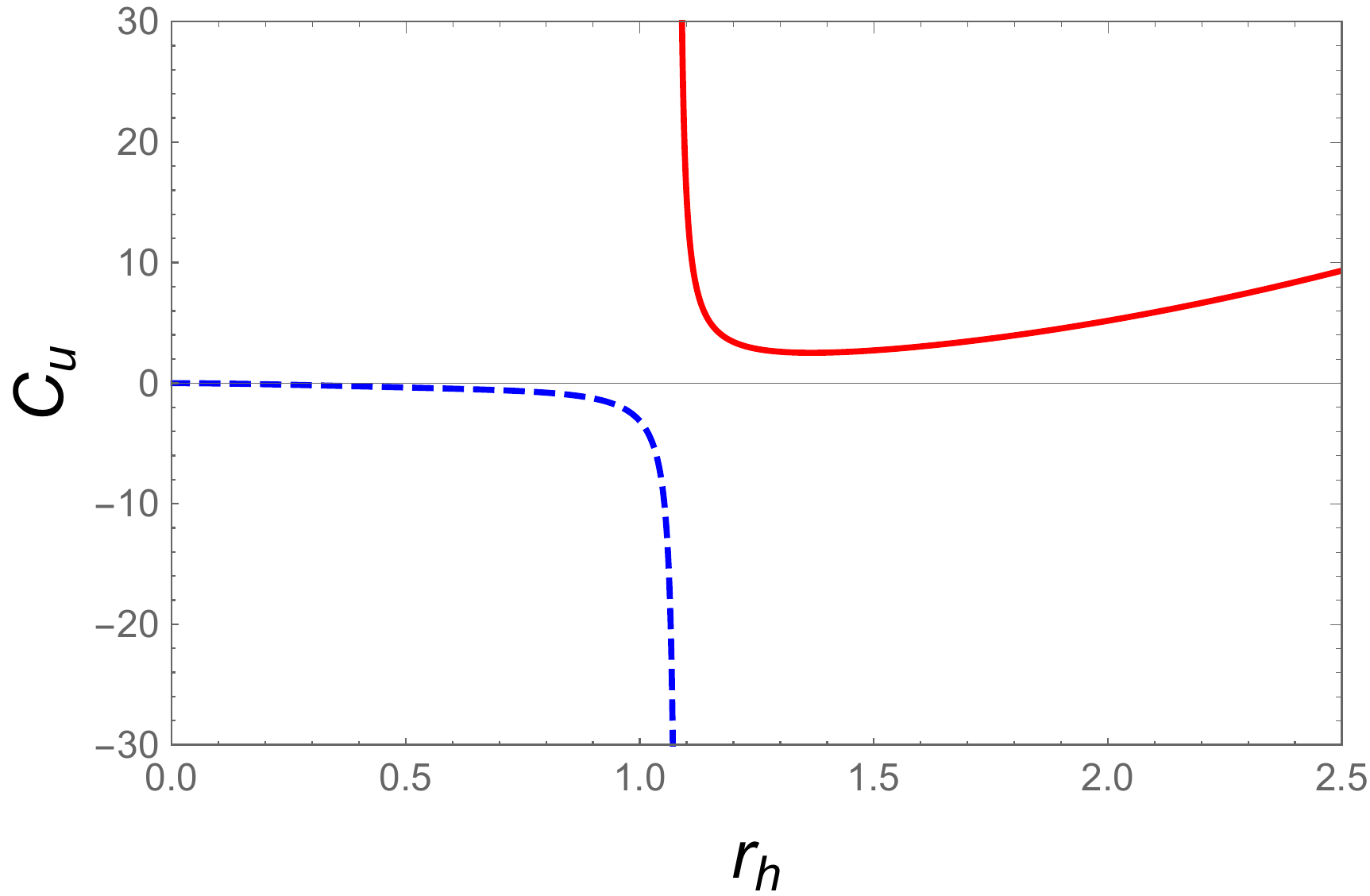}
\caption{Heat capacity $C_{u}$ of the the charged dRGT black string as a function of the horizon radius $r_h$ in the grand canonical ensemble,  with the parameters $\alpha_{m}=\alpha_{g}  =1$, $u  = 1$, $c_1  = 3$ and $c_0 = 4.5$.}
\label{m17}	
\end{center}
\end{figure}
In this regard, we can get the radial component of the vector field
\begin{equation}
\phi^{r_h}=\frac{1}{4 \tau} \alpha_g\left(-u^2 \tau-4 \pi r_h+\tau\left(c_0+\left(-2 c_1+3 r_h\right) r_h\right) \alpha_m^2\right).
\end{equation}
and the inverse temperature is
\begin{equation}
\tau=\beta=-\frac{4 \pi r_h}{u^2-c_0 \alpha_m^2+2 c_1 \alpha_m^2 r_h-3 \alpha_m^2 r_h^2}.
\end{equation}
It can also be found that, within the grand canonical ensemble, the inverse temperature of the charged dRGT black string  vanishes in two distinct limiting regimes. On the one hand, the inverse temperature approaches zero as the event horizon radius approaches its minimum allowed value. On the other hand, in the limit of an infinitely large horizon radius, the inverse temperature likewise decreases to zero. These results indicate that the system exhibits a clear asymptotic suppression of the inverse temperature in both the small-scale and large-scale limits. Accordingly, the corresponding asymptotic behavior may be constrained in the following form:
\begin{equation}\label{cmy1}
\beta\left(r_m\right)=0  \quad \text { and } \quad \beta(\infty)=0.
\end{equation}
\begin{figure}[h]
\begin{center}
\includegraphics[width=0.4\textwidth]{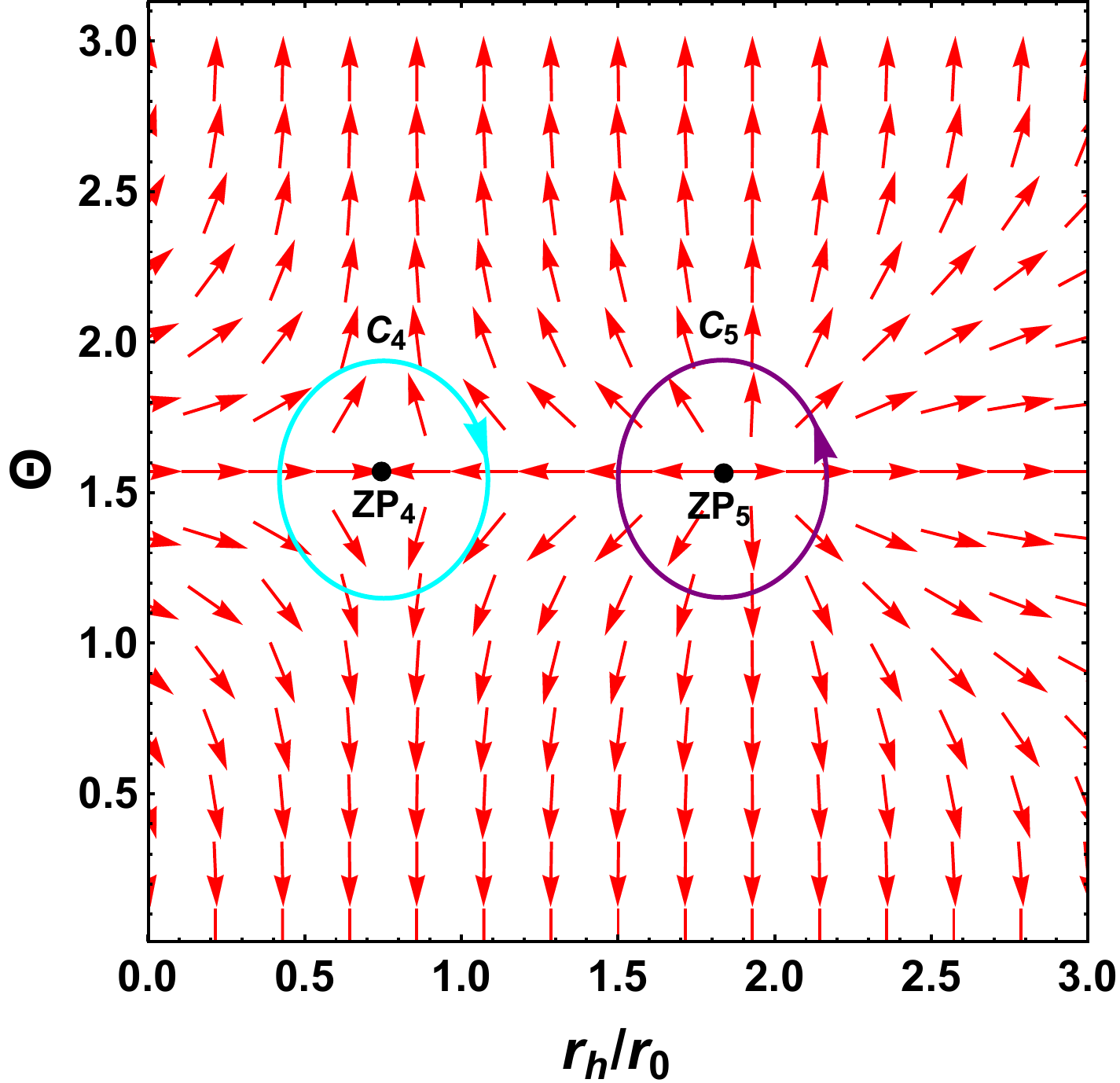}
\caption{The unit vector field $n$ on the $r_{h}-\Theta$ plane for the charged dRGT black string in the context of the  grand canonical ensemble with the parameter  $\beta / r_0 = 6$. }
\label{m8}	
\end{center}
\end{figure}
The function curve between the radius of the horizon and the inverse temperature  is shown in Fig.(\ref{m7}). One can identify a blue annihilation point $(\frac{d^2 \beta }{d r_{h}^2}<0)$ situated at the critical location $(9.3758,1.0801)$. For inverse temperatures below this critical threshold, there are both unstable branches of the small black string and stable branches of the large black string. In contrast, once the inversion temperature becomes greater than the critical value, no black string configuration can exist. In a manner similar to the analysis presented in the previous subsection, the local thermodynamic stability of the charged dRGT black string can be effectively described using its heat capacity. Within the grand canonical ensemble framework, the heat capacity of the black string is given by
\begin{equation}
\begin{aligned}
C_{u}&=T\left(\frac{\partial S}{\partial T}\right)_{u}\\
& =\frac{\pi r_h^2 \alpha_g\left(-u^2+\left(c_0+r_h\left(-2 c_1+3 r_h\right)\right) \alpha_m^2\right)}{u^2-\left(c_0-3 r_h^2\right) \alpha_m^2}.
\end{aligned}
\end{equation}
As shown in Fig. (\ref{m17}), the heat capacity displays a single divergence point, which precisely matches the horizon radius associated with the annihilation point shown in Fig. (\ref{m7}). This divergence divides the charged dRGT black string solutions into two separate branches: a thermodynamically unstable branch with negative heat capacity $( r_{h} \subset ( 0, 1.0801) )$, and a stable branch with positive heat capacity $( r_{h} \subset (1.0801, \infty ) )$. Such a result is fully consistent with the stability behavior inferred from the inverse temperature profile in Fig. (\ref{m7}), offering an independent confirmation of the thermodynamic topological analysis. It is also worth emphasizing that this two-branch structure, representing a transition from instability to stability, closely resembles the thermodynamic characteristics of the Hawking-Page phase transition. Furthermore, Fig. (\ref{m8}) illustrates the distribution of zero points for the unit vector field. Under the condition that the inversion temperature parameter satisfies $\beta / r_0 = 6$, two zero points are generated at  $(0.7805, \frac{\pi}{2})$ and $(1.9177, \frac{\pi}{2})$,respectively. To further characterize the corresponding winding numbers, two closed contours, $(C_4)$, and $(C_5)$, are constructed around these zero points.  According to the deflection angle curve displayed in Fig. (\ref{m9}), the associated topological charges are obtained as $Q_{ZP_{4}}=-1$ and $Q_{ZP_{5}}=1$. We analyze the behavior of the vector field $\varphi$ along the boundary specified by Eq. (\ref{cmy1}). The vector field tends to orient rightward as $(r_{h}\rightarrow r_{m})$ or $(r_{h}\rightarrow \infty)$, while the corresponding directional patterns are summarized in Table (\ref{tb1}). Fig. (\ref{m10}) clearly depicts the directional evolution of the vector field along the selected contours.  
By examining the clockwise and counterclockwise orientations of the corresponding closed loops, the winding numbers associated with the zero points $(ZP_{4}$ and $ZP_{5})$ are obtained as $-1$ and $+1$. The universal thermodynamic behavior of the charged dRGT black string within the grand canonical ensemble can be summarized as follows: in the low-temperature limit $(\beta \rightarrow \infty)$, the system does not admit any black string configuration. Conversely, in the high-temperature regime $(\beta \rightarrow 0)$, two separate branches arise, namely an unstable small black string phase and a stable large black string phase. This indicates that the charged dRGT black string and the Schwarzschild-AdS black hole share identical topological structure properties, and both belong to the  $(W^{0-})$ class \cite{us1}.
\section{Discussions}\label{IIII}
\begin{figure}[h]
\begin{center}
\includegraphics[width=0.4\textwidth]{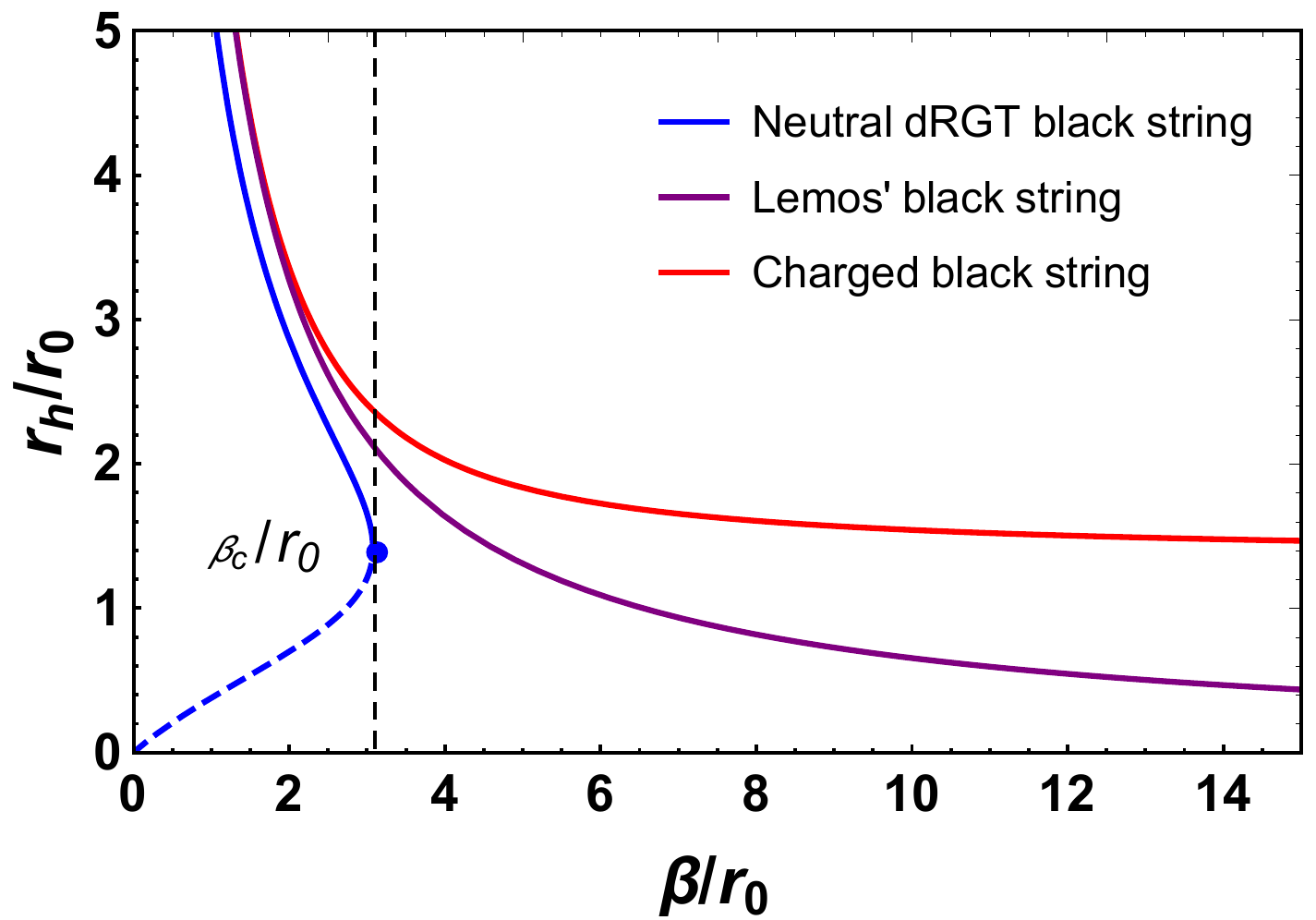}
\caption{In the $r_h-\beta$ plane, the zero points of the vector $\phi^{r_h}$. Neutral dRGT black string(Blue line) with parameters $\alpha_{m} / r_0 =\alpha_{g} / r_0 =1$,$c_1  / r_0 = 2.2$,$c_0  / r_0 = 6$ and $q  / r_0 =0$. Lemos' black string (Purple line) with parameters $\alpha_{m} / r_0 =\alpha_{g} / r_0 =0.8$,$c_1  / r_0 =c_0  / r_0 =q  / r_0 =0$. Charged black string (Red line) with parameters $\alpha_{m} / r_0 =\alpha_{g} / r_0 =0.8$,$c_1  / r_0 =c_0  / r_0 =0$,$q  / r_0 =1$. }
\label{m11}	
\end{center}
\end{figure}
To date, a systematic study on the universal thermodynamic topological classification of charged dRGT black strings has been performed in both the canonical and grand canonical ensemble frameworks. Our analysis reveals that the thermodynamic stability branches associated with these two ensembles possess markedly distinct topological structures. In particular, while the outermost large black string branch maintains thermodynamically stable behavior in both ensembles, the innermost small black string branch demonstrates fundamentally different stability properties: it is characterized as a thermodynamically unstable phase within the canonical ensemble, whereas in the grand canonical ensemble it instead belongs to a thermodynamically stable phase. These findings indicate that the choice of ensemble plays a crucial role in determining the thermodynamic topological structure and stability of the charged dRGT black string. Motivated by these results, we next proceed to discuss the topological classification of the charged dRGT black string in several limiting regimes.
\\
$(i)$ The neutral dRGT black string $(\gamma=0)$\\
Under the condition of ignoring the charge contribution,  the dependence of the inversion temperature on the horizon radius is illustrated in Fig. (\ref{m11}). One can observe that the asymptotic behavior of the system closely resembles that of the charged black string in the grand canonical ensemble. The distinction between these two cases mainly manifests through the displacement of the annihilation point $(3.0760, 1.4142)$. This behavior indicates that the neutral dRGT black string preserves the same thermodynamic topological classification as the charged black string within the grand canonical ensemble framework, and therefore their associated topological structures can both be classified into the $(W^{0-})$ -topological class.\\
$(ii)$ The charged black string $(c_1=c_0=0)$\\
When the effect of dRGT gravity on the black string is neglected, Fig. (\ref{m11}) indicates that the charged black string maintains thermodynamic stability in both the high-temperature and low-temperature asymptotic regimes. Meanwhile, the innermost small black string state together with the outermost black string state also exhibit stable thermodynamic behavior. This implies that the system is stable not only in the local thermodynamic sense but also in terms of its global thermodynamic configuration. Consequently, the associated vector field possesses a single zero point with winding number $(+1)$ (see References \cite{us2,us3,us4,us5,us6,us7,us8} for further details). It should be noted that, despite the pronounced difference in local thermodynamic stability between the charged black string and the charged dRGT black string in the canonical ensemble-where the latter develops an intermediate thermodynamically unstable branch-their temperature asymptotics remain consistent. As the event horizon radius increases, the winding numbers corresponding to the zero-point distribution follows the pattern $[+,(-,+),(-,+)]$. Hence, from the perspective of universal thermodynamic topological classification, both the charged black string and the charged dRGT black string in the canonical ensemble share an identical topological structure\cite{us1}, and they are classified within the $(W^{1+})$-topological category.\\
$(iii)$ The lemos' black string $(c_1=c_0=\gamma=0)$\\
Under the limiting conditions $\alpha_m=\alpha_g$ and $c_0=c_1=\gamma=0$, the charged dRGT black string degenerates to the solution of the Lemos' black string. As illustrated in Fig. (\ref{m11}), both the Lemos' black string and the charged black string display identical asymptotic behaviors of temperature, while the system remains thermodynamically stable in both the high- and low- temperature regimes. Consequently, the Lemos' black string and the charged RN-AdS black hole possess the same universal thermodynamic topological structure, and can both be classified into the  $(W^{1+})$ topological class. More importantly, our analysis indicates that, once the dRGT massive gravity correction is neglected, the inclusion of electric charge does not modify the thermodynamic topology of the black string configuration. In contrast, within the dRGT massive gravity framework, the existence of charge can substantially alter the topological structure of the black string. This result indicates that the combined effects of the dRGT massive gravity background and the electric charge exert a pronounced impact on the thermodynamic topological structure of black strings, thereby highlighting the crucial role played by dRGT massive gravity corrections in the black string thermodynamic topology.
\begin{widetext}
\begin{table*}[ht]
    \centering
    \begin{tabular}{|l|c|c|c|c|c||c|}
        \hline
        \textbf{W}&\textbf{Black hole solutions }& \textbf{Innermost}& \textbf{Outermost} & \textbf{Low $T$} & \textbf{High $T$} & \textbf{DP}  \\
        \hline
        $W^{1+}$&  Lemos' black string  & \emph{Stable} &\emph{Stable} & Stable small & Stable large & 0  \\\hline
        $W^{1+}$& Charged black string  & \emph{Stable} &\emph{Stable} & Stable small & Stable large & 0 \\\hline
        $W^{0-}$& Neutral dRGT black string & \emph{Unstable} & \emph{Stable}& No &Unstable small + stable large & 1AP \\\hline
        $W^{1+}$& Charged dRGT black string in CE  & \emph{Stable} &\emph{Stable} & Stable small & Stable large & 1GP+1AP \\\hline
         $W^{0-}$&Charged dRGT black string in GCE & \emph{Unstable} & \emph{Stable}& No &Unstable small + stable large & 1AP
        \\
        \hline
    \end{tabular}
    \caption{The universal thermodynamic topological classifications of black strings, where GP and AP denote the generation point and annihilation point, respectively, and CE and GCE denote the canonical and grand canonical ensemble, respectively.}
   \label{tablet11}
\end{table*}
\end{widetext}
\section{Conclusions}\label{IIV}
\label{Conc}
In this work, we have explored the the universal thermodynamic topological properties of the charged dRGT black string in both the canonical and grand canonical ensemble frameworks. In addition, the thermodynamic topology and stability properties of the charged dRGT black string were further explored under three particular parameter limiting cases. The thermodynamic behaviors associated with the innermost small black string branch and the outermost large black string branch, together with their stability characteristics in the high- and low-temperature limits, are summarized in Table (\ref{tablet11}). The main conclusions of this paper are as follows : $(1)$ The ensemble choice can significantly modify the thermodynamic topological structure of charged dRGT black strings. In particular, the innermost small black string branch shows different stability behavior in the canonical and grand canonical ensembles, resulting in a change of the topological classification. Accordingly,  the first-order small / large black hole phase transition and the Hawking-Page phase transition are found in the two ensembles, respectively. $(2)$ The dRGT massive gravity correction plays a key role in the topological classification of black string thermodynamics. It is found that the coupling between the gravitational background and the charge will significantly affect the thermodynamic topological properties of the black string. Therefore, the dRGT massive gravity cannot be regarded as a perturbation correction of the topological structure, but an important factor determining the thermodynamic topological configuration. The above results further expand the topological framework of black string thermodynamics and provide a new theoretical perspective for understanding the thermodynamic stability of black holes, topological phase structure and deep physical mechanism in gravity theory.
\acknowledgments
We acknowledge the anonymous referees for their valuable comments on improving our paper. This work is Supported by the National Natural Science Foundation of China (Grant No. 12247101 and 12405081), the Fundamental Research Funds for the Central Universities (Grant No. lzujbky-2025-jdzx07), the Natural Science Foundation of Gansu Province (No.25JRRA799), and the '111 Center' under Grant No. B20063, the project of Young Scientific and Technical Talents Development of Education Department of Guizhou Province (No.~[2024] 79).

\end{document}